\documentclass[conference]{IEEEtran}
\IEEEoverridecommandlockouts

\usepackage{cite}
\usepackage{amsmath,amssymb,amsfonts}
\usepackage{algorithmic}
\usepackage{graphicx}
\usepackage{textcomp}
\usepackage{xcolor}
\usepackage{booktabs}

\usepackage{hyperref}
\usepackage{url}
\usepackage{cite}

\usepackage{amssymb}
\usepackage{mathtools, nccmath}
\usepackage{multirow}  
\def\BibTeX{{\rm B\kern-.05em{\sc i\kern-.025em b}\kern-.08em
    T\kern-.1667em\lower.7ex\hbox{E}\kern-.125emX}}
\begin{document}

\title{BANC: Towards Efficient Binaural Audio Neural Codec for Overlapping Speech}
\author{\IEEEauthorblockN{1\textsuperscript{st} Anton Ratnarajah}
\IEEEauthorblockA{\textit{University of Maryland, College Park, MD, USA} }
\and
\IEEEauthorblockN{2\textsuperscript{nd} Shi-Xiong Zhang}
\IEEEauthorblockA{\textit{Tencent AI Lab, Bellevue, WA, USA} }
\and
\IEEEauthorblockN{3\textsuperscript{rd} Dong Yu}
\IEEEauthorblockA{\textit{Tencent AI Lab, Bellevue, WA, USA}}

}

\maketitle

\begin{abstract}
We introduce BANC, a neural binaural audio codec designed for efficient speech compression in single and two-speaker scenarios while preserving the spatial location information of each speaker. Our key contributions are as follows: 1) The ability of our proposed model to compress and decode overlapping speech. 2) A novel architecture that compresses speech content and spatial cues separately, ensuring the preservation of each speaker's spatial context after decoding. 3) BANC's proficiency in reducing the bandwidth required for compressing binaural speech by 48\% compared to compressing individual binaural channels. In our evaluation, we employed speech enhancement, room acoustics, and perceptual metrics to assess the accuracy of BANC's clean speech and spatial cue estimates.
\end{abstract}


\section{Introduction}

\label{sec:intro}
Neural audio codecs (NACs) are used to compress audio signals into codes to reduce the amount of data transmitted or stored. Existing NACs can be categorized into hybrid approaches and end-to-end NACs. Hybrid NACs combine traditional signal processing-based audio coding methods with neural speech synthesis architectures~\cite{hybrid_codec1,hybrid_codec2,hybrid_codec3}. Recently, data-driven end-to-end NAC architectures have been proposed~\cite{soundstream,encodec,hificodec,AudioDec}, substantially improving the quality of reconstructed audio. The end-to-end approach does not assume the nature of the input audio training dataset, allowing it to generalize to any audio content. Most NACs are composed of encoder, quantizer, and decoder modules. Current NACs are designed to compress single-channel or stereo audio signals, and their decoder modules are typically optimized for single-speaker scenarios~\cite{AudioDec}. Two major drawbacks are evident in prior NAC designs:
1) They predominantly focus on single-channel audio processing.
2) They assume the presence of only one primary speaker within the audio.
To address these limitations, our paper introduces a novel neural spatial audio codec model, aptly named BANC. This model efficiently compresses binaural speech, accommodating both single-speaker and two-speaker scenarios with different spatial locations.


A key distinction between single-channel and binaural audio is the latter's encapsulation of spatial localization, in addition to clean speech content ($S_C[t]$)~\cite{ITD}. This spatial context manifests in various acoustic facets, such as early reflections, late reverberations, diffractions, as well as interaural discrepancies like interaural time difference (ITD) and interaural level difference (ILD) between microphones. Mathematically, these intricacies can be captured by the impulse response (IR) function. Consequently, for a mono-speaker scenario, one can decompose the clean speech content $S_C[t]$ and the binaural acoustic effect $BIR[t]$ from the binaural speech $S_B[t]$ as:
{\small
\belowdisplayskip 0.8\belowdisplayshortskip
\begin{equation}\label{reverberant_speech} 
\begin{aligned}[b]
    S_B[t] = S_C[t] \circledast BIR[t].
\end{aligned}
\end{equation}
}

For overlapped binaural speech ($S_{OB}[t]$) with two speakers in two different spatial locations, we can separately decompose their clean speech contents ($S_{C1}[t]$, $S_{C2}[t]$) and their binaural IRs (BIRs) ($BIR_1[t]$, $BIR_2[t]$) as follows:
{\small
\belowdisplayskip 0.4\belowdisplayshortskip
\begin{equation}\label{overlap_speech}
\begin{aligned}[b]
    S_{OB}[t] =  (S_{C1}[t] \circledast BIR_1[t]) + (S_{C2}[t] \circledast BIR_2[t]) .
\end{aligned}
\end{equation}
}


\textbf{Main Contribution: } We present a pioneering NAC architecture optimized for binaural overlapped speech from two speakers, crucially retaining each speaker's spatial context. This architecture is illustrated in Fig.~\ref{architecture}. In contrast to the existing AudioDec model\cite{AudioDec}, our key contributions are: 
1) Expansion of previous neural codecs from single to binaural audios.
2) Enhanced capability for compressing and decoding overlapping speech.
3) A novel strategy of separately compressing speech content and spatial cues, ensuring post-decoding preservation of spatial context for each speaker.
4) Achieving a high compression rate for binaural audio. Specifically, our model can reconstruct a 48 kHz binaural speech signal from two distinct speakers using only 12.6 kbps bandwidth, a feat surpassing AudioDec which operates at 24 kbps for binaural speech and results in a remarkable 52\% reduction in spatial errors, respectively. Moreover, BANC demonstrates superiority over both Opus\cite{opus} and Encodec~\cite{encodec} in quantitative and qualitative analysis when tested under comparable bandwidths.
We provide reconstructed speech samples, spectrograms and source code for future research at \url{https://anton-jeran.github.io/MAD/}. 
    

\section{Related Work}
\label{sec:related}

\textbf{Traditional audio codecs:} Linear predictive coding-based audio codecs~\cite{traditional_codec1,traditional_codec2} and model-based audio codecs~\cite{traditional_codec3} have been proposed in the past for speech coding, but their quality is limited. Among traditional methods, Opus~\cite{opus} and EVS~\cite{EVS} are state-of-the-art traditional audio codec architectures, and they can support different bitrates and sampling rates at high coding efficiency in real-time. 

\textbf{Neural audio codecs (NAC):}  End-to-end data-driven architectures have been proposed to encode mono and stereo audio with impressive performance~\cite{soundstream,encodec,hificodec}. Encodec~\cite{encodec} compresses stereo audio by processing the left and right channels separately, which leads to inefficient compression as the same speech content is encoded twice in both channels. Our BANC model reduces bandwidth by encoding speech content only once and is optimized to efficiently compress overlapping speech while preserving each speaker's speech content and spatial acoustic features.

\textbf{Speech dereverbation and RIR Estimation:} Recently, NAC-based architectures have been proposed for audio-visual speech enhancement~\cite{codec_enhancement}. Similarly, in our work, we decode clean speech from binaural speech. Generative architectures have also been proposed to estimate IR from given spatial information~\cite{fast-rir,mesh2ir}. Encoder-decoder architectures have shown promising results in estimating IR from reverberant speech~\cite{estimate1,estimate2}. We propose a neural codec to estimate the IR of a one-second duration from binaural speech.
\section{Binaural Audio Neural Codec}
\label{multi-audio}
We propose BANC to compress binaural speech $S_B(x)$ with a sampling rate of 48 kHz. Similar to typical NAC~\cite{soundstream,AudioDec}, our model consists of an encoder, projector, quantizer and decoder modules. We propose simple and complex decoder architecture for single-speaker and two-speaker scenarios, respectively. Our proposed encoder architecture is the same for single-speaker and two-speaker cases. We adapt the projector and quantizer from the AudioDec~\cite{AudioDec}.

\begin{figure*}[t] 
	\centering
	\includegraphics[width=0.75\linewidth]{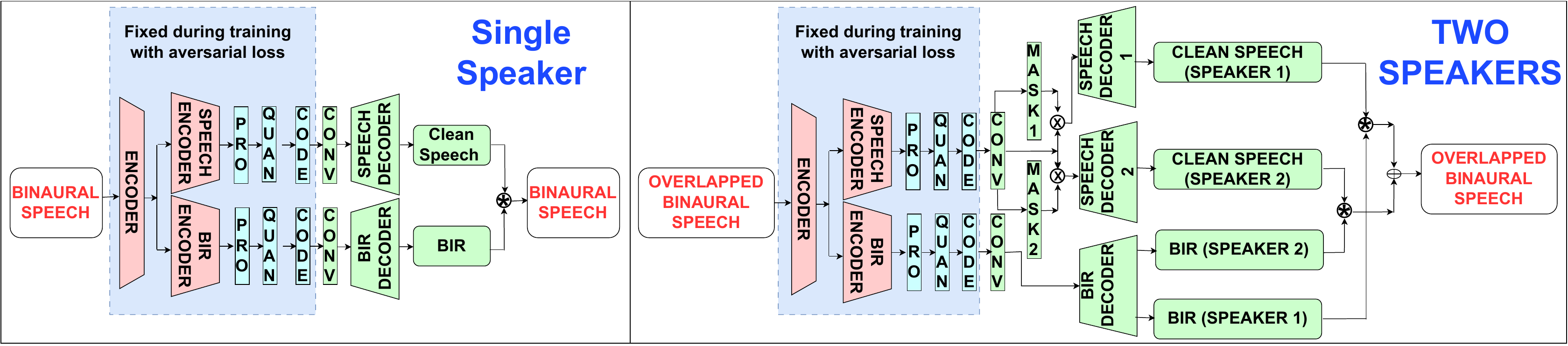}
	\caption{Our proposed NAC configured for single-speaker and two-speaker two-spatial overlapped binaural speech. In \S~\ref{multi-audio}, we describe the details of our architecture and training paradigm. We train end-to-end networks with metric loss (Eq.~\ref{metric_loss}) for 200K iterations. Then, we freeze the blocks shown in blue and train the rest of the network with the metric and adversarial loss (Eq.~\ref{adversarial_loss}) for an additional 500k iterations (single-speaker) and 160k iterations (two-speaker).} 
	\label{architecture}
\end{figure*}
\subsection{Encoder Architecture}
\label{encoder_archi}
We pass the binaural speech through a common encoder, consisting of a 1D convolutional layer (CONV) with a kernel size (K) of 3, stride (S) of 1, input channels (IC) of 2, and output channels (OC) of 2. The output from the common encoder is then passed to both the speech encoder and the binaural impulse response (BIR) encoder. The speech encoder follows the same architecture as AudioDec~\cite{AudioDec} and SoundStream~\cite{soundstream}. The speech encoder starts with a CONV layer (K = 7, S = 1, IC = 2, OC = 16), followed by convolution blocks $C_{B1}$. Each $C_{B1}$ contains three residual units (RU) with dilated CONV layers (with dilation rates of 1, 3, and 9), followed by a CONV layer (K = 2 * S, S = S, IC = IC, OC = 2 * IC). We have 5 $C_{B1}$ blocks with strides (S) of (2, 2, 3, 5, 5), resulting in a downsampling factor of 300 for the speech content. \\

Our IR encoder is inspired by the IR estimator network~\cite{estimate1}. The IR encoder contains three CONV blocks, $C_{B2}$. Each $C_{B2}$ consists of a CONV layer followed by batch normalization (BN) and leaky ReLU. The first $C_{B2}$ does not include BN. The three $C_{B2}$ blocks have OC = (128, 256, 512), K = (96001, 41, 41), S = (1500, 2, 2), and padding (P) = (48000, 20, 20). This architecture significantly downsamples the IR content by a factor of 6000. All the CONV layers are causal to enable real-time operation. The outputs of both the speech encoder and IR encoder are projected into multi-dimensional space separately and quantized into codes using projector and quantizer modules, as proposed in AudioDec.

\subsection{Decoder Architecture}
\label{decoder_archi}
We propose two different architectures for the single-speaker and two-speaker scenarios as follows:\\

\textbf{Single speaker: } We propose a speech decoder architecture to decode clean speech and an IR decoder to decode BIR. We reconstruct the binaural speech from the estimated clean speech and IR using Eq.~\ref{reverberant_speech}. Both the speech and IR decoders are adaptations of the SoundStream decoder. Before feeding inputs into the decoder modules, we pass the code through a CONV layer (IC = 64, OC = 512, K = 7, S = 1). \\

The speech decoder consists of 5 CONV blocks $C_{B3}$ with S = (5, 5, 3, 2, 2), followed by a CONV layer with OC = 1, K = 7, and S = 1. Each $C_{B3}$ contains transposed convolutional layers (IC = IC, OC = 0.5 * IC, K = 2 * S, S), followed by three residual units (RU) layers similar to those in the encoder. The IR decoder has a similar network structure as the speech decoder, except for the number of $C_{B3}$ blocks. The IR decoder contains 6 $C_{B3}$ blocks with S = (5, 5, 5, 4, 3, 2), and the final CONV layer has two output channels. We reconstruct two-second clean speech and one-second BIR at a sampling rate of 48 kHz. \\

\textbf{Two speakers:} For the binaural scenario, we replicate the speech decoder from the single-speaker scenario twice to decode the clean speech of two speakers separately. We perform speech separation within the decoder to ensure the network preserves the speech content of each individual speaker in the code. Instead of directly passing the output $C$ from the CONV layer, we learn the representation of each speaker, $S_{i}$, by learning a mask vector $M_{i} \in [0,1]$. Similar to Conv-TasNet~\cite{Conv-TasNet}, $S_{i}$ is calculated by performing element-wise multiplication of $C$ and $M_{i}$. We then pass $S_{i}$ to the speech decoder modules to estimate the clean speech of each speaker. The same IR decoder from the single-speaker network is used, but with double the number of channels in each layer. Fig.~\ref{architecture} illustrates our model for two-speaker binaural speech. \\

\textbf{Bandwidth:} AudioDec allocates 80 bits per frame at a sampling rate of 48,000 Hz and a stride factor of 300, resulting in a bandwidth of 25.6 kbps for binaural speech (2 channels) calculated as 2 × 80 × 48,000 / 300. In contrast, our proposed BANC method compresses the clean speech signal by a factor of 300 and the BIR by 6000, leading to a significantly reduced bit rate of 13.44 kbps, computed as (80 × 48,000 / 300) + (80 × 48,000 / 6000). \\

\subsection{Training Objective}
We adapt the training paradigm proposed in AudioDec. First, we train the end-to-end network with the metric loss for 200k iterations. Then, we replace our speech decoders with HiFi-GAN~\cite{HiFi-GAN} vocoders and continue training with both the metric and adversarial losses for an additional 500k iterations, using HiFi-GAN-based multi-period and multi-scale discriminators~\cite{AudioDec}. In the multi-speaker scenario, after 200k iterations, we further train our end-to-end network with adversarial and metric losses for an additional 160k iterations. Let $B(x)$ and $\hat{B}(x)$ denote the input and reconstructed binaural speech, respectively. We denote the ground truth and reconstructed clean speech of speaker $i$ as $S_{i}(x)$ and $\hat{S}{i}_(x)$, respectively. $BIR_{i}(x)$ and $\hat{BIR}_{i}(x)$ represent their corresponding ground truth and reconstructed binaural impulse responses.\\

\textbf{Metric Loss: } 
We use the mel spectral loss (Eq.~\ref{mel_loss}) and spectrogram loss as our metric loss for clean and binaural speech. In Eq.~\ref{mel_loss}, $MEL$ denotes the extraction of the mel spectrogram. $\mathbb{E}$ denotes the expectation, and L1-norm and L2-norm are denoted by $||.||_1$ and $||.||_2$ respectively.

{\small
\begin{equation}\label{mel_loss}
\begin{aligned}[b]
    \mathcal{L}_{MEL}(x,\hat{x}) = \mathbb{E}[||MEL(x) - MEL(\hat{x})||_{1}].
\end{aligned}
\end{equation}
}
 
For spectrogram loss, we calculate the mean square difference of the log magnitude of the ground truth speech spectrogram ($M_{spec}(x)$) and estimated speech spectrogram ($M_{spec}(\hat{x})$) (Eq.~\ref{spec_mag_loss}).
 {\small
\begin{equation}\label{spec_mag_loss}
\begin{aligned}[b]
    \mathcal{L}_{MAG}(x,\hat{x}) = \mathbb{E}[||M_{spec}(x) - M_{spec}(\hat{x})||_{2}^2].
\end{aligned}
\end{equation}
}

 We calculated time-domain mean square error (MSE) between ground truth and estimated BIRs (Eq.~\ref{bir_loss}) as our metric loss for estimated BIRs as follows:

 {
 \small
\begin{equation}\label{bir_loss}
\begin{aligned}[b]
    \mathcal{L}_{IR}(b,\hat{b}) = \mathbb{E}[||b - \hat{b}||_{2}^2] .
\end{aligned}
\end{equation}
}

Our total metric loss is defined as follows:
 {
  \small
\begin{equation}\label{metric_loss}
\begin{aligned}[b]
    \mathcal{L}_{MET}(x) = (\mathcal{L}_{MEL}(B(x),\hat{B}(x)) +  \mathcal{L}_{MAG}(B(x),\hat{B}(x))  ) 
    \\ + \sum_{i=1}^{M_S}  (\mathcal{L}_{MEL}(S_{i}(x),\hat{S}_{i}(x))  +  \mathcal{L}_{MAG}(S_{i}(x),\hat{S}_{i}(x))   
    \\+ \mathcal{L}_{IR}(BIR_{i}(x),\hat{BIR}_{i}(x)) ),
\end{aligned}
\end{equation}
}

where $M_S$ is the total number of speakers in the speech.\\ 

\textbf{Adversarial Loss: } We train two HiFi-GAN discriminators for binaural and clean speech by optimizing the following objective function:
 {
 \small
\begin{equation}\label{discriminator_loss}
\begin{aligned}[b]
    \mathcal{L}_{D}(x) = \mathbb{E}[\max{(0,1-D_{B}(B(x)))} + \max{(0,1+D_{B}(\hat{B}(x))}
    \\ + \sum_{i=1}^{M_S}(\max{(0,1-D_{S}({S}_{i}(x)))} + \max{(0,1+D_{S}(\hat{S}_{i}(x))}))], 
\end{aligned}
\end{equation}
}

where $D_{B}$ and $D_{S}$ are the discriminators of binaural speech and clean speech respectively. We train our BANC with the following adversarial loss.
 {
 \small
\begin{equation}\label{adversarial_loss}
\begin{aligned}[b]
    \mathcal{L}_{ADV} = \mathbb{E}[\max{(0,1-D_{B}(\hat{B}(x))} 
     + \sum_{i=1}^{M_S} \max{(0,1-D_{S}(\hat{S}_{i}(x))})] .
\end{aligned}
\end{equation}
}

In addition to the $\mathcal{L}_{MET}(x)$ and $\mathcal{L}_{ADV}(x)$, we train our network with $\mathcal{L}_{VQ}$~\cite{VQ} applied to the VQ codebook. Our overall generator loss $\mathcal{L}_{GEN}(x)$ is as follows:
{
\begin{equation}\label{generator_loss}
\begin{aligned}[b]
    \mathcal{L}_{GEN}(x) = \mathcal{L}_{MET}(x) + \lambda_{ADV} \mathcal{L}_{ADV} + \lambda_{VQ} \mathcal{L}_{VQ},
\end{aligned}
\end{equation}
} 
where $\lambda_{ADV}$ and $\lambda_{VQ}$ are the weights.
\section{Experiments}
\label{sec:experiments}
\textbf{Dataset:} We generate binaural speech datasets for both single-speaker and two-speaker scenarios using clean speech from the Valentine dataset~\cite{valentini, VCTK} and simulated BIRs from pygsound~\cite{pygsound}. We simulated 50k BIRs using pygsound, which were then randomly convolved with the clean speech corpus using Eq.\ref{reverberant_speech} for single-speaker and Eq.\ref{overlap_speech} for two-speaker scenarios. We split the simulated dataset into 33,975 training samples, 750 validation samples, and 752 test samples.

\noindent\textbf{Baselines:} Opus is a widely used audio codec in Zoom, Microsoft Teams, Google Meet, and YouTube, and it was standardized by the IETF in 2012. We use Opus as our baseline traditional audio codec. We also compare our approach with the state-of-the-art NAC for two-channel audio (Encodec)\cite{encodec}. HiFi-Codec\cite{hificodec} and AudioDec~\cite{AudioDec} only support single-channel speech compression. Therefore, we separately compressed the left and right channels using HiFi-Codec and AudioDec. The pre-trained AudioDec model available on their official GitHub was trained only on clean speech. For a fair comparison, we trained AudioDec using our dataset. AudioDec is an improvised version of SoundStream~\cite{soundstream} for speech coding. More details about our baseline can be found in Table~\ref{table1}.

\noindent\textbf{Ablation:} We evaluated three variations of BANC to select the best model for the single-speaker case. We assess the benefit of the HiFi-GAN vocoder by training our network for 700k iterations using our simple speech decoder described in \S~\ref{decoder_archi} (BANC-V1). In AudioDec, only mel spectral loss is used as a metric loss. Therefore, we trained the network without Eq.\ref{spec_mag_loss} to evaluate the benefits of spectrogram loss (Eq.\ref{spec_mag_loss}) (BANC-V2). BANC is trained using our proposed approach described in \S~\ref{multi-audio}. Due to computational complexity, we do not use the HiFi-GAN vocoder for our two-speaker model.

\noindent\textbf{Evaluation Metrics:} 
We evaluate our model by measuring the clean speech estimation quality using the widely used speech enhancement metric STOI~\cite{stoi} and BIR estimation quality using a set of BIR acoustic parameters. Reverberation time ($T_{60}$), direct-to-reverberant ratio (DRR), early-decay time (EDT), and early-to-late index (CTE) are commonly used acoustic parameters to measure IRs~\cite{ir-gan,ts-rir}. We calculate the mean absolute difference between the estimated and ground truth BIR acoustic parameters.

We measure our model's ability to preserve interaural time difference (ITD) and interaural level difference (ILD) in reconstructed binaural speech. As proposed in prior work~\cite{itd1}, we use the generalized cross-correlation phase transform (GCC-PHAT) algorithm~\cite{gcc_phat} to calculate the ITD error (Eq.~\ref{itd_loss}) between the left and right channels of the ground truth speech ($B^L$, $B^R$) and the reconstructed speech ($\hat{B}^L$, $\hat{B}^R$).
 {
 \small
\begin{equation}\label{itd_loss}
\begin{aligned}[b]
    \mathbf{E_{ITD}} = \mathbb{E}[|ITD(B^L,B^R) - ITD(\hat{B}^L,\hat{B}^R)|] .
\end{aligned}
\end{equation}
}

We define the ILD error for left channel ($\mathbf{E_{ILDL}}$) and right channel ($\mathbf{E_{ILDR}}$) as follows:
{
 \small
\begin{equation}\label{ild_loss1}
\begin{aligned}[b]
    \mathbf{E_{ILDL}} = \mathbb{E}[|20\log_{10}\frac{||\hat{B}^L||_2^2}{||B^L||_2^2}|] ,  \mathbf{E_{ILDR}} = \mathbb{E}[|20\log_{10}\frac{||\hat{B}^R||_2^2}{||B^R||_2^2}|] . 
\end{aligned}
\end{equation}
\vspace{-0.3cm}
}
\vspace{-0.3cm}
\begin{table}[h!]
    \setlength{\tabcolsep}{1.5pt}
	\renewcommand{\arraystretch}{0.75} 
	\caption{The baselines used for the comparison show that BANC significantly compresses the binaural speech. BANC can reduce the bandwidth of AudioDec by up to 47.5\% for binaural speech. Opus~\cite{opus} and HiFi-Codec~\cite{hificodec} does not report their compression and bandwidth respectively.}
	\label{table1}
	\centering
 \resizebox{0.7\columnwidth}{!}{
	\begin{tabular}{@{}lccccc@{}}	
		\toprule
		\textbf{Method} & \textbf{Compression} &\textbf{Bandwidth}  & \textbf{Sampling Rate}  \\
		\midrule
        Opus-12~\cite{opus}& - & 12 kbps  & 48 kHz\\
        Opus-24~\cite{opus}& - & 24 kbps  & 48 kHz\\
        HiFi-Codec-320~\cite{hificodec}& 320x & -  & 24 kHz\\
	HiFi-Codec-240~\cite{hificodec}& 240x & -  & 24 kHz  \\
        Encodec-12~\cite{encodec}& 256x & 12 kbps & 48 kHz  \\
        Encodec-48~\cite{encodec}& 64x & 48 kbps & 48 kHz  \\
        AudioDec~\cite{AudioDec}& 300x & 25.6 kbps & 48 kHz\\
        \textbf{BANC (Ours)} & \textbf{3150x} & \textbf{13.44 kbps} & \textbf{48 kHz} \\
	\midrule
		
	\end{tabular}
 }
	\vspace{-0.5cm}
\end{table}
\begin{table}[h!]
    \setlength{\tabcolsep}{1.5pt}
	\renewcommand{\arraystretch}{0.75} 
	\caption{Interaural time difference error ($\mathbf{E_{ITD}}$) and interaural level difference errors ($\mathbf{E_{ILDL}}$, $\mathbf{E_{ILDR}}$) of the final reconstructed binaural speech are reported for the baselines (Table~\ref{table1}) and different variations of our model (\S~\ref{sec:experiments}). We also report the STOI of the intermediate estimated clean speech from our approach. Additionally, we compare our single and two-speaker models (Fig.~\ref{architecture}).
}
	\label{table2}
	\centering
     \resizebox{0.7\columnwidth}{!}{
	\begin{tabular}{@{}llcccc@{}}	
		\toprule
        \textbf{Speakers} & 	\textbf{Method} & \textbf{$\mathbf{E_{ITD}}\downarrow$} &\textbf{$\mathbf{E_{ILDL}}\downarrow$}  & \textbf{$\mathbf{E_{ILDR}}\downarrow$}    & \textbf{STOI$\uparrow$}\\
                       
	\midrule
        Single & Opus-12                             & 30.7 ms          & 1.28  &  1.28  & - \\
        Single & Opus-24                             & 25.4 ms          & 1.04  &  1.07  & - \\
        Single & HiFi-Codec-320                      & 36.0 ms          & 1.21  &  1.24  & - \\
        Single & HiFi-Codec-240                      & 37.8 ms          & 1.07  &  1.08  & - \\
        Single & Encodec-12                          & 33.5 ms          & 0.88  &  0.91  & - \\
        Single & Encodec-48                          & 34.0 ms          & \textbf{0.56}  &  \textbf{0.57}  & - \\
        Single & AudioDec                            & 33.7 ms          & 0.87           &  0.88           & - \\
        Single & BANC-V1                   & 29.0 ms          & 1.20           &  1.36           & 0.71 \\
        Single & BANC-V2                   & 20.7 ms          & 1.33           &  1.41           & 0.71 \\
        \textbf{Single} & \textbf{BANC}      & \textbf{16.0 ms} & 0.75           &  0.72           & \textbf{0.84} \\
        \midrule
        Two & Opus-12                             & 27.3 ms          & 1.01  &  1.00  & - \\
        Two & Opus-24                             & 31.2 ms          & 0.71  &  0.79  & - \\
        Two & Encodec-12                          & 33.5 ms          & 0.83  &  0.81  & - \\
        Two & Encodec-48                          & 36.4 ms          & \textbf{0.50}  &  \textbf{0.50}  & - \\
        \textbf{Two} & \textbf{BANC}      & \textbf{21.9 ms} & 0.82           &  0.77           & \textbf{0.72} \\
        \midrule

		
	\end{tabular}
 }
\end{table}
\vspace{-0.5cm}
\begin{table}[!h]
    \setlength{\tabcolsep}{1.5pt}
	\renewcommand{\arraystretch}{0.75} 
	\caption{BIR estimation error for our approach in single-speaker and two-speaker scenarios. Training binaural speech with spectrogram loss significantly improves BIR estimation indirectly. The BIR estimation of our network in the two-speaker scenario is comparable to that in the single-speaker scenario when using a simple speech decoder (BANC-V1). For the two-speaker case, we report the average error across both speakers. }
	\label{table3}
	\centering
  \resizebox{0.7\columnwidth}{!}{
	\begin{tabular}{@{}ccccccc@{}}	
		\toprule
        \textbf{Speakers}& \textbf{Channel}  & \textbf{Method}  & \textbf{$\mathbf{T_{60}}\downarrow$} & \textbf{DRR$\downarrow$} & \textbf{EDT$\downarrow$} & \textbf{CTE$\downarrow$}  \\
            &   & & \textbf{(ms)} & \textbf{(dB)} & \textbf{(dB)} & \textbf{(ms)}  \\
		\midrule
         Single                 &Left          & BANC-V1             & 25.3          & 2.79          & 86.7          & 2.23  \\
         Single                 &Left          & BANC-V2             & \textbf{20.9} & 2.21          & 67.0          & 1.44 \\
        \textbf{Single}         &\textbf{Left} & \textbf{BANC (ours)}& 22.7          & \textbf{1.08} &\textbf{39.4}  & \textbf{0.79}   \\
        \midrule
        \textbf{Two}     &\textbf{Left} & \textbf{BANC (ours)}& \textbf{25.2} & \textbf{3.41} &\textbf{80.1}  & \textbf{2.52}   \\
        \midrule
        \midrule
         Single                 &Right          & BANC-V1             & 23.8          & 2.84          & 84.7           & 2.09  \\
         Single                 &Right          & BANC-V2             & \textbf{21.4} & 2.35          & 64.9           & 1.33 \\
        \textbf{Single}         &\textbf{Right} & \textbf{BANC (ours)}& 23.0          &\textbf{1.05}  & \textbf{35.0}  & \textbf{0.77}  \\
        \midrule
        \textbf{Two}     &\textbf{Right} & \textbf{BANC (ours)}& \textbf{25.6} & \textbf{3.30} &\textbf{83.3}   & \textbf{2.09}   \\
        \midrule
        \midrule
		
		
	\end{tabular}
 }
\end{table}

\noindent\textbf{Results:}  Table~\ref{table2} presents the ITD and ILD errors of the reconstructed binaural speech from different baselines and our approach. Our approach achieves the lowest ITD error for both single-speaker and two-speaker cases. Additionally, it outperforms in terms of ILD errors ($\mathbf{E_{ILDL}},~\mathbf{E_{ILDR}}$) compared to all baselines, except for Encodec-48. However, Encodec-48 requires four times more bandwidth, has a compression rate approximately 50 times lower than our approach, and is only suitable for non-streamable usage. For a fair comparison, we compared our model with Encodec-12 and found that our approach outperforms it by 18\% and 3\% for single-speaker and two-speaker cases, respectively. We also compared three variations of our single-speaker model, and observed that replacing the simple speech decoder with a HiFi-GAN vocoder improves the ITD error by 29\%, while adding spectrogram loss enhances clean speech estimation quality (STOI) by 15\%.

Table~\ref{table3} shows the BIR estimation error for our approach. We observed that improving the binaural speech estimation quality using the HiFi-GAN vocoder and spectrogram loss indirectly contributed to better BIR estimation, reducing the overall error for $T_{60}$, DRR, EDT, and CTE by 6.9\%, 62.1\%, 56.6\%, and 64\%, respectively, in the single-speaker scenario. The performance of our two-speaker model is comparable to that of our single-speaker model BANC-V1. 

\noindent\textbf{Perceptual Evaluation:} We randomly selected two different single-speaker reverberant speech signals, compressed and decoded them using different audio codecs, and asked 26 participants from Amazon Mechanical Turk to rate the quality compared to the ground truth speech on a scale of 1 to 100. The average completion time of the survey is around 10 minutes. The average scores were 69.06, 64.83, 69.81, 70.71 and 72.87 for Opus-24, HiFiCodec-24, Encodec-24, AudioDec, and our BANC model, respectively. These results demonstrate that the audio reconstructed by BANC outperforms prior codecs.



		

\section{Conclusion and Future Work}
\label{sec:conclusion}
We propose BANC, a novel binaural NAC for single-speaker speech and two-speaker spatially overlapped speech. Our approach outperforms traditional methods and NACs with similar bandwidth, preserving binaural acoustic effects by up to 52\%. We introduce a novel technique to compress speech content and acoustic effects separately, demonstrating that our method can reduce the bandwidth for compressing binaural speech by 48\% compared to compressing each channel individually using AudioDec. Given the complexity of the scenario and the need for headphone-based evaluation, we have tested our approach on binaural two-speaker spatially overlapped speech. In the future, we aim to extend it to compress and decode overlapped speech from multiple speakers in arbitrary locations.

\bibliographystyle{IEEEtran}
\bibliography{IEEEexample}
\end{document}